\documentclass{aa}
\usepackage{graphicx}

\begin{document}

\sloppypar

%
   \title{Distribution of the Galactic Bulge emission at $|b|>2^\circ$ according to the RXTE Galactic Center scans}

   \author{M. Revnivtsev \inst{1,2}}

   \offprints{mikej@mpa-garching.mpg.de}

   \institute{Max-Planck-Institute f\"ur Astrophysik,
              Karl-Schwarzschild-Str. 1, D-85740 Garching bei M\"unchen,
              Germany,
        \and   
              Space Research Institute, Russian Academy of Sciences,
              Profsoyuznaya 84/32, 117810 Moscow, Russia
            }
  \date{}

        \authorrunning{Revnivtsev et al.}
        \titlerunning{Distribution of the Galactic Bulge emission at $|b|>2^\circ$}
 
   \abstract{We present an analysis of the Galactic bulge emission 
observed by the RXTE/PCA during a set of scans over the Galactic Center field, 
performed in 1999-2001. The
total exposure time of these observations is close to 700 ksec.
We construct the distribution of Galactic ridge emission intensity 
and spectral parameters up to Galactic latitudes $b\sim -10^\circ;+9^\circ$. 
We show that the intensity distribution of the ridge emission at 
$|b|>2^\circ$ could be well described by an exponential model with e-folding
width $b_0\sim 3.3^\circ$. Best-fit spectral parameters do not show
statistically significant changes over Galactic latitude
   }

   \maketitle

%

\section{Introduction}
Galactic ridge emission was detected by the rocket experiment in 1972 (
\cite{bleach}). It was noted that there exists an excess in X-ray emission 
($>$1.5 keV) around the Galactic plane with an extent about 2-4$^\circ$.
Since then, various satellites were used to study this emission,
 HEAO1 (\cite{worrall}), EXOSAT (\cite{warwick}), Tenma (\cite{koyama86}),  
GINGA (\cite{yamasaki}), ASCA (\cite{kaneda97}), RXTE (\cite{valinia}) and
Chandra (\cite{ebisawa}). 

The observation of the ASCA and Chandra observatories seem to rule out the hypothesis of
dominant dim point source contributions to the observed ridge emission 
(see \cite{ebisawa}). Therefore it is believed that it has a diffuse origin.

The discovery by the Tenma satellite of strong 6.7 keV line emission in the spectrum of the
Galactic plane made it possible to suggest that the bulk of this Galactic ridge emission 
in the energy range 1--10 keV is due to an optically thin plasma of 
temperature of a few keV.  Accurate measurements of the ridge spectrum 
made by ASCA also revealed lines from some other elements, Mg, Si, S, 
Ar, which also supports the thermal origin of the emission. However, this
hypothesis also encounters serious problems. One of the most general
problems is connected with the fact that the deduced parameters of the 
optically thin plasma implies that it is impossible to bound such plasma 
within Galactic plane (see e.g. \cite{townes89}) and also it is very 
hard to provide enough energy for such plasma. There are also 
serious problems with the approximation of the energy spectra of the ridge 
emission within the framework of its thermal origin (\cite{tanaka00}). 
These complications
together with the detection of the presumably non-thermal tail in the spectrum of the
Galactic ridge emission (see e.g. \cite{yamasaki}, \cite{skibo}, 
\cite{valinia}) gives rise to the additional interpretation, in which 
the X-ray line emission was considered to originate through charge-exchange 
interactions of low-energy cosmic ray heavy ions (e.g. \cite{tanaka99}, 
\cite{tanaka00}), while the hard power-law tail appears as a result of 
nonthermal bremsstrahlung emission of cosmic ray electrons and protons 
(e.g. \cite{dogiel}).

For the understanding of the origin of the Galactic ridge emission it is
important to know the distribution of its flux and parameters over the 
Galaxy. Such a study has been previously carried out using different satellites 
(e.g. HEAO1, \cite{worrall}; \cite{iwan82}, GINGA; \cite{yamasaki}, RXTE \cite{valinia}), but now we could for the first time use relatively uniform
coverage of the central 10$^\circ$ degrees. This became possible because of
the large campaign of RXTE Galactic Center scans, organized by
the RXTE team. In this paper we analyze public 
data of RXTE Galactic center scans from March 1999 till July 2001.

\section{Analysis}

For our analysis we used approximately 150 sequences  of 
$\sim$3200 Galactic Center observations (each set of Galactic center 
scan sequences usually consists of 22 individual observations) performed 
from March 1999 till July 2002. 
The total exposure of all these observations is approximately 700 ksec.
 We divided these data
into two parts, depending on high voltage epoch of PCA, which
determines the energy response of the instrument: 
Apr.1999-- May 13. 2000 (Epoch4) and May 14, 2000-July 2001 (Epoch 5).

For the data analysis we used a set of standard procedures
of the LHEASOFT 5.2 package. In order to increase our sensitivity for 
photons with energy $>$10 keV we used all three layers of the PCA.
As we are interested in the measurement of low fluxes
we used the ``L7\_240'' model for the PCA background estimation. This model 
includes an instrumental background as well as the Cosmic X-ray
background (CXB) term. The influence of interstellar absorption in 
the direction of Galactic Center could be important at low $b$. However,
in the subsequent analysis we will restrict ourself to $|b|>2^\circ$ 
where the interstellar absorption is negligible in 
our bandpass ($3-20$ keV). Under these condition,
extinction of CXB in the interstellar medium is estimated to
be at most 10 
1990), thus can be ignored.

According to the
latest calibration information the systematic uncertainty of the
flux obtained with the help of the background model used  is considered
to be  within $\sim$1--1.5\%  
(see RXTE GOF web page. http://heasarc.gsfc.nasa.gov/docs/xte/xhp\_proc\_analysis.html). Therefore
in our analysis we included 1.5\% (of the background count rate 
in the considered energy range) uncertainty in the measurements.

\begin{table}
\caption{List of point sources, areas around which were excluded from the analysis\label{sourcelist}}
\tabcolsep=0.4cm
\tiny
\begin{tabular}{ccl}
\hline
l&b&Source\\
\hline
 -9.812&  -0.861& XTEJ1723-376 \\ 
 -8.527&  -0.548& EXO1722-363 \\ 
 -7.943&   2.746& 4U1711-339 \\ 
 -7.831&  -0.272& AXJ172642-3540 \\ 
 -6.596&  -3.386& XTEJ1743-363 \\ 
 -6.469&  -5.005& 4U1746-371 \\ 
 -6.398&  -0.827& 1RXSJ173251.1-344728 \\ 
 -5.698&  -0.150& 4U1728-34 \\ 
 -5.604&   1.257& 1RXSJ172635.1-325842 \\ 
 -5.159&  -0.158& MXB1730-33 \\ 
 -4.978&   3.346& 1RXSJ172006.1-311702 \\ 
 -4.611&  -8.151& SAXJ1808.4-3658 \\ 
 -3.681&   2.298& TERZAN2 \\ 
 -3.633&   6.931& XTEJ1710-281 \\ 
 -3.378&   0.219& AXJ173628-3141 \\ 
 -3.184&  -2.976& SL1746-331 \\ 
 -2.875&  -1.607& H1741-322 \\ 
 -2.530&   7.911& XTEJ1709-267 \\ 
 -2.442&   0.989& XB1732-304 \\ 
 -1.892&   0.520& XTEJ1739-302 \\ 
 -1.154&   1.393& GRS1734-292 \\ 
 -0.880&  -0.101& 1E1740.7-2942 \\ 
 -0.860&  -2.908& XTEJ1755-312 \\ 
 -0.744&  -0.911& SLX1744-300/299 \\ 
 -0.441&  -0.389& 2S1742-294 \\ 
 -0.155&  -3.126& XTEJ1757-306 \\ 
 -0.135&   8.741& 1RXPJ171236-2414.7 \\ 
  0.535&   9.278& 2RXPJ171220.5-232345 \\ 
  0.667&  -0.036& SGR B2 \\ 
  0.676&  -0.222& XTEJ1748-288 \\ 
  0.785&   2.398& SLX1735-269 \\ 
  1.074&   3.655& KS1731-260 \\ 
  1.119&  -1.028& X1749-285 \\ 
  1.530& -11.371& R1832-330 \\ 
  1.937&   4.795& GX1+4 \\ 
  2.294&   0.794& GX3+1 \\ 
  2.788&  -7.914& 4U1820-30 \\ 
  2.862&  -0.680& 1RXSJ175454.2-264941 \\ 
  3.840&   1.463& HA1745-248 \\ 
  4.508&  -1.362& GRS1758-258 \\ 
  4.765&   0.608& 2E1751.1-2431 \\ 
  4.790&   3.316& XTEJ1744-230 \\ 
  5.077&  -1.019& GX5-1 \\ 
  6.141&  -1.904& 2S1803-245 \\ 
  6.381&  -0.120& E1757.5-2330 \\ 
  6.756&  -4.798& SAXJ1819.3-2525 \\ 
  7.554&   0.853& XTEJ1759-220 \\ 
  7.729&   3.802& NGC6440 \\ 
  8.151&  -0.712& SAXJ1806.5-2215 \\ 
  9.077&   1.154& GX9+1 \\ 
  9.275&  -6.081& GS1826-238 \\ 
  9.996&  -0.141& AXJ180816-2021 \\ 
 11.068&  -0.627& 1RXPJ181217-1939.4 \\ 
\hline
\end{tabular}
\end{table}

\begin{figure}
\includegraphics[width=0.5\textwidth,bb=130 150 470 630,clip]{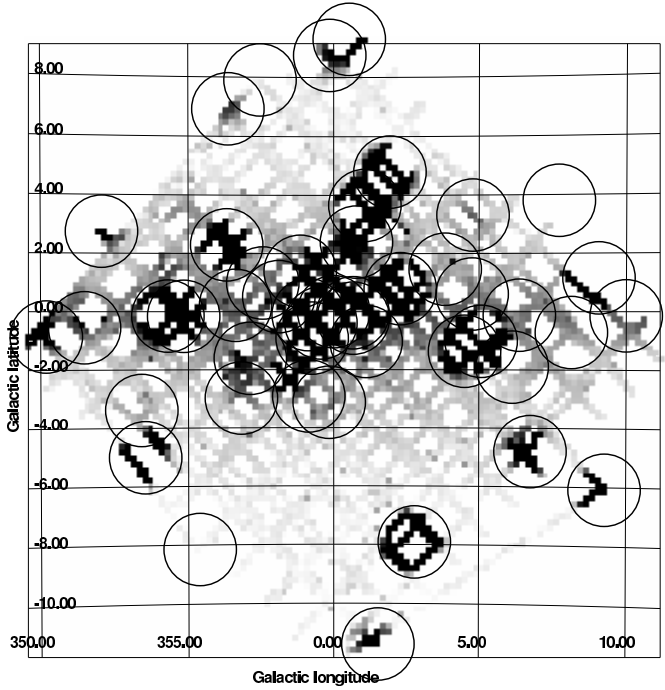}
\caption{Map of the Galactic Center field, reconstructed from the 
RXTE/PCA scans in 1999-2000. Circles represent regions where the contribution of
point sources dominates (see text). \label{maps}}
\end{figure}

\begin{figure}
\includegraphics[width=0.5\textwidth]{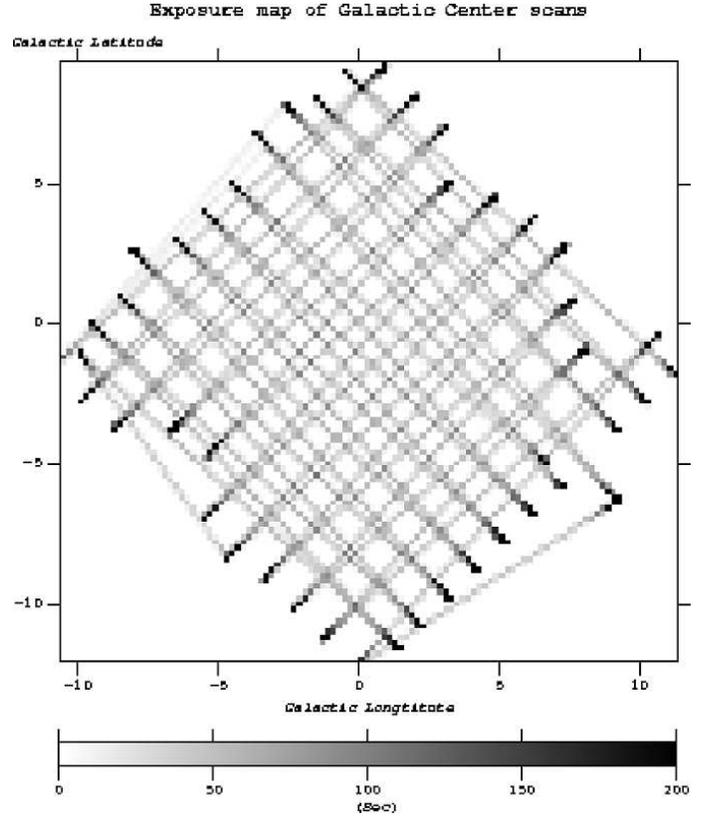}
\caption{Exposure map of the Galactic scan observations during period 
March 1999 - May 2000 (high voltage epoch4)}
\label{exposuremap}
\end{figure}

The map of the Galactic center region, reconstructed from scans performed
during epoch 4 (March 1999 - May 2000) is presented in Fig. \ref{maps}.
The map represents the flux measured by PCA in the direction to
       which the center of the PCA field of view was pointed.
 Any point source on the map contributes 
to a sky region around it in accordance with the response of the RXTE/PCA 
collimator.  Combining information from available X-ray catalogs of 
bright sources
(\cite{wood84}, \cite{hmxb}, \cite{lmxb}, \cite{rass}, \cite{sugizaki01}) 
with the analysis of the obtained map, we have selected a set of sources that
 can significantly contribute to the observed flux in the region
of the scans. A list of the selected sources is presented 
in Table \ref{sourcelist}. Note that in our analysis the identification
 of sources 
in crowded regions, like in the immediate vicinity of the Galactic Center, 
is quite complicated and therefore might not be exact.

In order to separate the contribution of bright point sources from the 
Galactic
 ridge  emission 
we excluded areas with a radius of 1.35$^{\circ}$ around them -- such a 
radius ensures that even the brightest sources (such as GX5-1) 
do not contribute to the surrounding points more than a few cnts/s/PCU to the
surrounding points. 

Upper limits on the unaccounted sources within the field of the scans could
be estimated as $\sim$1--2 cnts/s/PCU ($\sim$0.5--1 mCrab) at Galactic 
latitudes $|b|\ga1-2^\circ$. Assuming a Crab-like spectrum of sources 
this upper limit corresponds to a flux $\sim 10^{-11}$ erg/s/cm$^2$. 

The effective field of view of the RXTE/PCA spectrometer is about $\approx 1$ 
deg$^2$.  According to the
luminosity function of the Galactic X-ray sources, measured e.g. by 
ASCA (\cite{ueda99}, \cite{sugizaki01}), RXTE/ASM (\cite{grimm}), 
CHANDRA (\cite{ebisawa}) at Galactic Latitudes 
$|b|<0.5$ the density of sources with a flux higher than $\sim10^{-11}$ 
erg/s/cm$^2$ (i.e. compatible with our rejection limit) is of the order 
of 0.1 deg$^{-1}$ and this value drops with increasing $|b|$ 
(e.g. \cite{grimm}). The
contribution of weaker point sources to the ridge emission does not exceed 
approximately $10\%$ (e.g. \cite{ebisawa}). Therefore, due to our 
limited sensitivity to weak point sources the obtained brightness profiles
and spectra could slightly suffer from their influence and an additional
'noise' of $\approx$10\% could appear.

\section{Results}

\subsection{Brightness distribution of the ridge emission}

After the rejection of regions affected by point sources
 we have hardly any data 
within $|b|\la 2^\circ$. Besides, a large number of weak point sources 
(see e.g. \cite{sugizaki01}) in this region could strongly contaminate 
the ridge emission observed by the RXTE/PCA. Therefore we will not try 
to study the ridge emission at these latitudes in detail.

During Epoch4 and Epoch5 the PCA detectors had 
significantly different response functions, therefore we will analyze data 
obtained during these periods separately.

\begin{figure}
\includegraphics[width=\columnwidth]{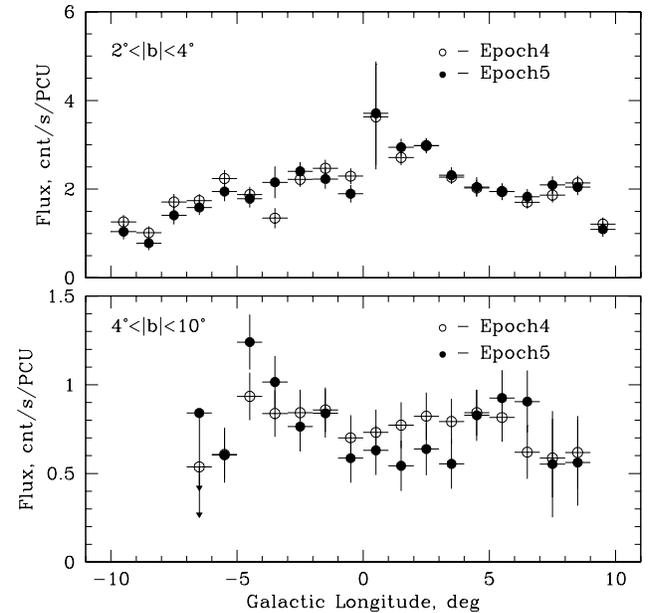}
\caption{Distribution of the 3-20 keV intensity of the Galactic ridge emission with longitude. The regions over which the 
brightness is averaged are 
$2 < |b| < 4$ for the upper plot,
and $4 <|b|<10$ for the lower plot}
\label{long_profiles}
\end{figure}

Measured profiles at latitudes higher than 2$^\circ$ show 
a quite weak dependence on longitude, see Fig. \ref{long_profiles}.

\begin{figure*}
\includegraphics[width=0.5\textwidth]{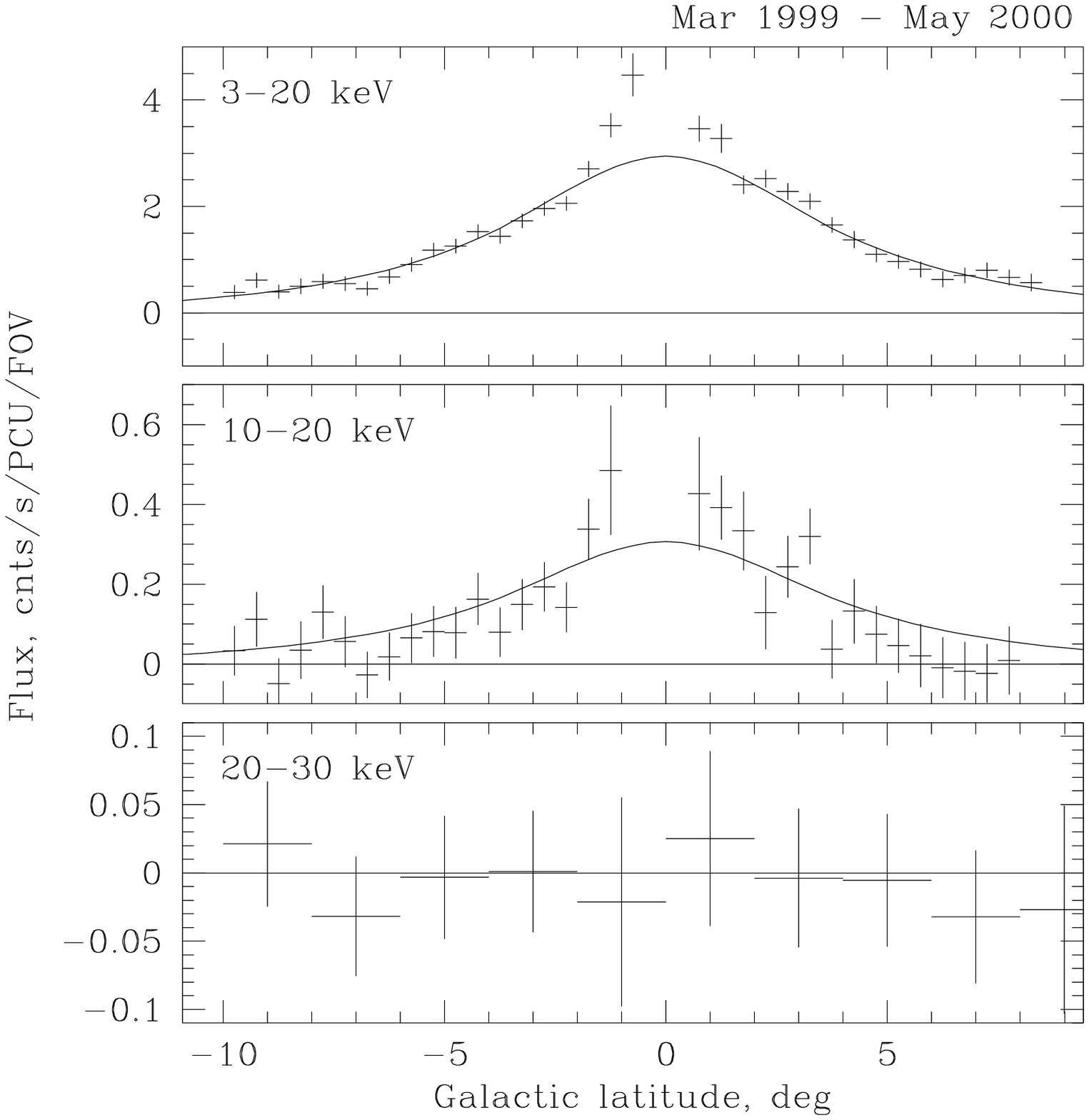}
\includegraphics[width=0.5\textwidth]{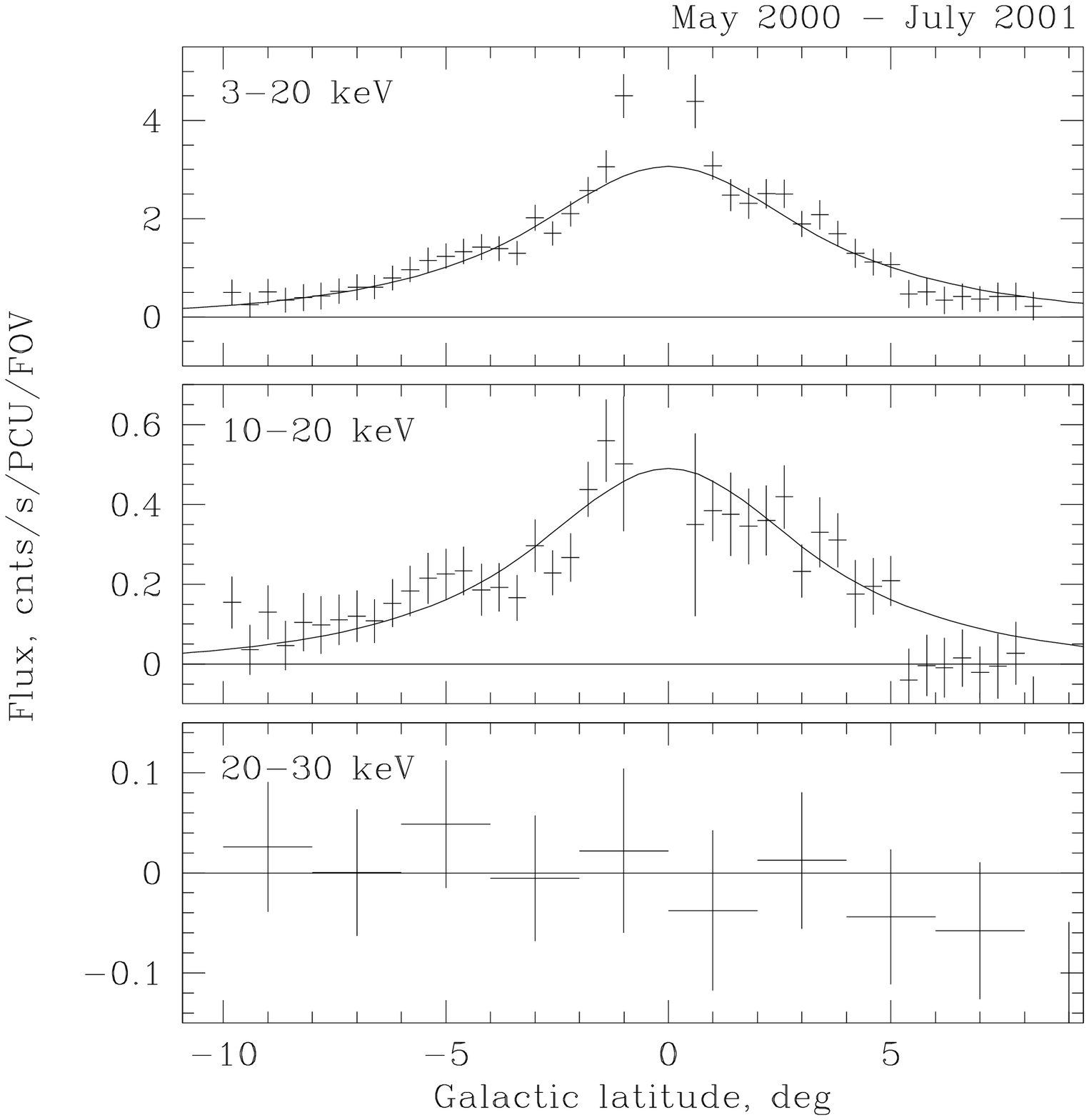}
\caption{The profile of the brightness of the Galactic ridge 
emission with latitude in three energy bands constructed using 
March 1999-May 2000 data (Epoch4, left) and May 2000-July 2001 data
(Epoch5, right). The solid line shows the convolution 
of the exponential model with the response of the RXTE/PCA collimator. 
There is a clear indication of the presence of an additional 
component within $|b|<1-2^\circ$. Taking into account the observed 
spectral shape, the 1 cnts/s/PCU corresponds approximately to 
$1.2\cdot 10^{-11}$ erg/s/cm$^2$ in the 3-20 keV energy band, 
approximately to $2.3\cdot 10^{-11}$ erg/s/cm$^2$ in the 10--20 keV energy 
band and $\sim 10^{-10}$ erg/s/cm$^2$ in the 20-30 keV energy band. Here 
and below the bars represent $1-\sigma$ errors}
\label{profiles}
\end{figure*}

We constructed profiles
of the brightness of the Galactic ridge emission, averaged over all longitudes 
($l$) in our scan field. The intensity profiles in three energy bands 
are presented in Fig. \ref{profiles}. 

The latitude distribution of the 
Galactic ridge emission at $|b|>2^\circ$ can be well described by an
exponential model of the form $I={\rm Norm}\cdot \exp(-|b|/b_0)$.
The PCA collimator collects X-rays from approximately
1 sq.deg, and therefore the measured profile of the ridge emission
is in reality a convolution of the sky distribution with the response of 
the PCA collimator. Therefore in our approximation of the observed profiles
we folded the model profile with the response of the PCA collimator.

The measured profiles of the X-ray intensity of the ridge
emission and the best fit exponential models (folded with the 
collimator response) are presented in Fig. \ref{profiles}. 
Note that there are indications that an additional
narrow component is present at low latitudes, $b<$1--2$^{\circ}$. A similar
narrow component was also found by Valinia \& Marshall (1998). But in
our case strong
contamination by point sources in the Galactic plane does not 
allow us to study this component in detail. Our obtained profiles
at $|b|<4^\circ$ are consistent with the results of Valinia \& Marshall 
(1998), but their model of the spatial variation of the 
intensity of the ridge emission
(a Gaussian with FWHM long Galactic latitude $\sim4.8^\circ$) can
no longer describe the profile of the diffuse emission at higher 
latitudes. One should use the exponential model instead.

The parameters of the approximations of the observed profiles at $|b|>2^\circ$ in the two energy 
bands 3--20 keV and 10--20 keV are presented in Table \ref{prof_param}.
The presence of the Galactic ridge emission in 20--30 keV energy band
of PCA is not statistically significant in the analyzed data.

\begin{table}
\caption{Approximation of observed brightness profiles of Galactic ridge emission in two energy bands}
\label{prof_param}
\tabcolsep=0.1cm

\begin{tabular}{l|cc|cc}
Par.&\multicolumn{2}{c}{Epoch4}&\multicolumn{2}{c}{Epoch5}\\
\hline
         & 3--20 keV& 10--20 keV& 3--20 keV&10--20 keV\\
\hline
$b_0$&$3.30\pm 0.58$&$3.27\pm1.28$&$3.25\pm 0.71$&$3.26\pm1.78$\\
Norm$^a$&$4.53\pm 0.42$&$0.45\pm0.09$&$4.73\pm0.52$&$0.68\pm0.18$\\
\hline
\end{tabular}
\begin{list}{}
\item $^a$ - normalization in counts/s/PCU. 1 count/s/PCU corresponds
approximately to $1.2\cdot 10^{-11}$, $2.3\cdot 10^{-11}$ and $10^{-10}$ erg/s/cm$^2$ in 3--20 keV, 10--20 keV and 20--30 keV energy bands correspondingly
\end{list}
\end{table}

\subsection{Approximation of the spectrum of the ridge emission}

The spectrum of the Galactic ridge emission is known to be quite rich, full of
different lines of different elements (see e.g. \cite{kaneda97}).
Unfortunately most of these lines lie in energy bands lower than 3-3.5 keV,
i.e. below our bandpass. In our energy range we can see only the blend of
Fe lines at energies around 6-7 keV. 

As was shown before (e.g. \cite{yamasaki}, \cite{valinia}), the 3--20 keV
 spectrum of the Galactic ridge emission could be relatively well described 
by a single power law
with a gaussian line at energy $\sim$6.7 keV.
the quite large exposure time of the collected data (approximately 200 ksec 
for each epoch after subtraction of contaminated regions) allows us to make
a spectral approximation of the observed ridge emission at different latitudes.
 Below we will use a power law with a gaussian line 
model for the approximation of the the ridge emission collected over
the whole scan field with $|b|>2^\circ$, as well as for emission collected 
over individual 1$^\circ$-width 
strips along the Galactic plane.

\begin{figure}
\includegraphics[width=0.5\textwidth]{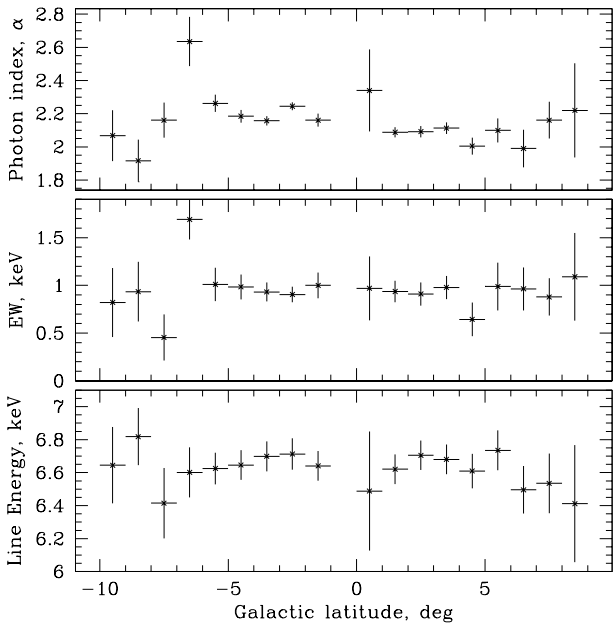}
\caption{Latitude distribution of parameters of spectral approximation of the Galactic ridge emission (combined data of Epoch 4 and 5)}
\label{params_profile}
\end{figure}

Best fit parameters of the approximation of the Galactic ridge spectrum
averaged over the whole scan field
with $|b|>2^\circ$ are presented in Table \ref{spectral_params}. Dependences of
best fit parameters on Galactic latitude are presented in Fig. \ref{params_profile}

\begin{figure}
\includegraphics[width=\columnwidth]{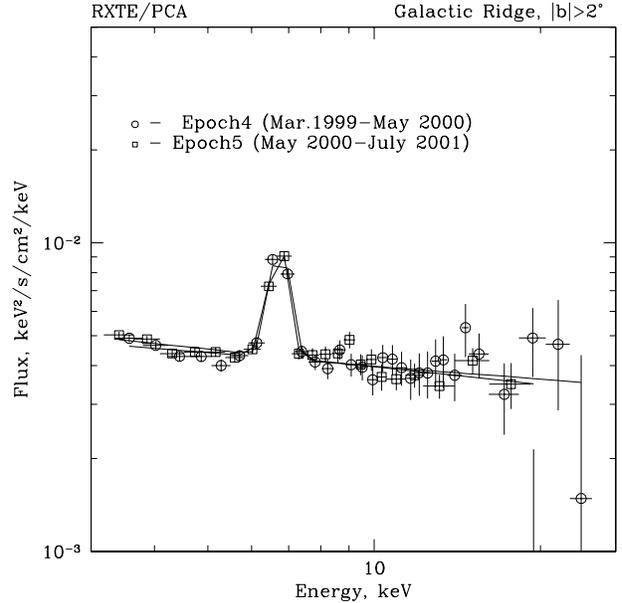}
\caption{PCA spectrum of the Galactic ridge emission collected over $|b|>2^\circ$}
\end{figure}

\begin{table}
\caption{Spectral approximation of data collected at $|b|>2^\circ$}
\label{spectral_params}

\begin{tabular}{l|c|c}

&Epoch 4 & Epoch5\\
\hline
Photon index, $\alpha$&$2.14\pm0.02$&$2.15\pm0.02$\\
Line energy, keV&$6.71\pm0.03$&$6.65\pm0.05$\\
Line width, keV&$<0.32$&$<0.39$\\
Line EW$^a$, eV&$882\pm60$&$810\pm70$\\
$\chi^2/dof$&33/45&51/45\\
\hline
\end{tabular}
\normalsize

$^a$ - equivalent width of the line
\end{table}

\section{Summary}

We analyzed the data of RXTE scans over Galactic center regions 
performed in 1999-2001. After subtraction of regions contaminated by 
bright pointed sources we constructed the intensity profile
of the Galactic ridge emission and its spectral parameters 
across the galactic plane within $|b|\la10^\circ$. We show
that the intensity profile at $|b|>2^\circ$ could be well described
by an exponential model ($\propto \exp(-|b|/b_0)$) with e-folding size $b_o\sim3.3^\circ$.
A spectral approximation of data collected over 1-deg strips along 
the galactic plane does not show statistically significant changes
of best fit parameters both of the continuum and of the Fe line. The averaged
spectrum of the ridge emission observed by the PCA could be approximated
by a power law with a slope $\alpha\sim2.15$ and a gaussian line at 
energy $\sim 6.7$ keV with
equivalent width $\sim 850$eV. These results are compatible with
previously published results of GINGA (\cite{yamasaki}) and RXTE 
(\cite{valinia}).

The origin of the Galactic ridge emission is still not known.
Recent results of ASCA and especially Chandra allowed us to exclude the 
possibility that the observed ridge emission is due to the integrated 
flux of faint point sources (\cite{ebisawa}) and we can conclude that 
it should have a truly diffuse origin. 
Different scenarios of the Galactic ridge emission production 
(e.g. magnetic reconnection, \cite{tanuma99}; supernova explosions, 
\cite{koyama86b}; cosmic rays \cite{tanaka99}, \cite{dogiel} and so on) could
in principle be distinguished by comparing different brightness profiles
predicted by models.

\begin{figure}
\includegraphics[width=\columnwidth]{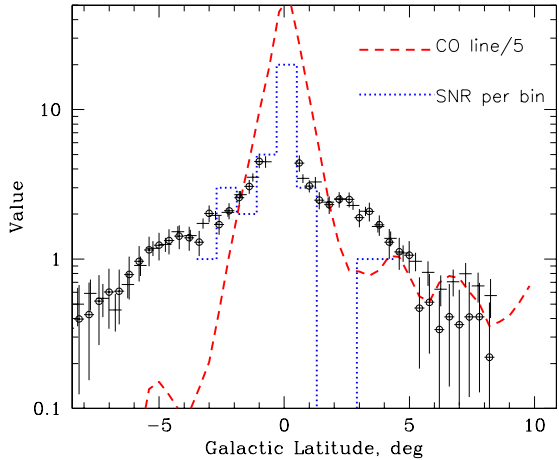}
\caption{Distribution of observed brightness of the Galactic ridge 
emission (crosses and open circles), the CO line intensity (short 
dashed line) and number of 
supernova remnants within 1 deg stripes along the galactic plane within 
field of scan of RXTE. Y-axis has units cnts/s/PCU for the observed 
Galactic ridge X-ray emission (3--20 keV),  K km s$^{-1}$ deg for CO line}
\label{difprofiles}
\end{figure}

Following Markevitch et al. (1993) and Yamauchi\& Koyama (1993) it is 
interesting to compare the observed profile of the ridge emission 
with the distribution of the CO line flux (e.g. \cite{coline}) as a tracer of 
molecular gas in the Galaxy and also with the density of SNR 
(\cite{green_snr}) as a tracer of supernova explosions. The constructed 
distributions within the field of scans of RXTE/PCA are presented in 
Fig.\ref{difprofiles}. It is clearly seen that CO line intensity drops 
much more abruptly than the intensity of the ridge emission, while the
distribution of SNR more closely resembles the profile of the ridge X-ray 
emission. Unfortunately,
poor statistics of SNR at high latitudes does not allow us to make
any solid conclusions. 

\begin{acknowledgements}
Author is grateful to Rashid Sunyaev, Marat Gilfanov and Eugene Churazov 
 for valuable discussions. 
This research has made use of data obtained through the High Energy
Astrophysics Science Archive Research Center Online Service,
provided by the NASA/Goddard Space Flight Center.
\end{acknowledgements}

\begin{thebibliography}{}
\bibitem[Bradt et al. 1993]{rxte} Bradt H., Rotshild R.,
Swank J. 1993, Astron. Astrophys. Suppl. Ser.  97, 355
\bibitem[Bleach et al. 1972]{bleach} Bleach R., Boldt E., Holt S. et al. 1972, ApJ, 174, 101L
\bibitem[Dame et al. 1987]{coline}Dame T. M., Ungerechts H., Cohen R. S. et al. 1987, ApJ, 322, 706
\bibitem[Dickey \& Lockman 1990]{nhmap} Dickey J., Lockman F. 1990, ARA\&A, 28, 215
\bibitem[Dogiel et al. 2002]{dogiel} Dogiel V., Schoenfelder V., Strong A. 2002, A\&A, 382, 730 
\bibitem[Ebisawa et al. 2001]{ebisawa} Ebisawa K., Maeda Y., Kaneda H. et al. 2001, Science, 293, 1633
\bibitem[Green 2001]{green_snr}Green D. 2001, http://www.mrao.cam.ac.uk/surveys/snrs/snrs.data.html
\bibitem[Grimm et al. 2002]{grimm} Grimm H.-J., Gilfanov M., Sunyaev R. 2002, A\&A, 391, 923
\bibitem[Iwan et al. 1982]{iwan82}Iwan D., Marshall F., Boldt E. et al. 1982, ApJ, 260, 111
\bibitem[Kaneda et al. 1997]{kaneda97} Kaneda H., Makishima K., Yamauchi S. et al. 1997, 491, 638
\bibitem[Koyama et al. 1986a]{koyama86} Koyama K., Makishima K., Tanaka Y. et al. 1986a, PASJ, 38, 121
\bibitem[Koyama et al. 1986b]{koyama86b} Koyama K., Ikeuchi S., Tomisaka K. 1986b, PASJ, 38, 503
\bibitem[Liu et al. 2000]{hmxb}Liu Q. Z., van Paradijs J., van den Heuvel E. P. J. 2000, A\&AS, 147, 25
\bibitem[Liu et al. 2001]{lmxb} Lui Q. Z., van Paradijs J., van den Heuvel E. P. J. 2001, A\&A, 368, 1021
\bibitem[Markevitch et al. 1993]{}Markevitch M., Sunyaev R., Pavlinsky M. 1993, Nature, 364, 40
\bibitem[Skibo et al. 1997]{skibo} Skibo J., Johnson W., Kurfess J. et al. 1997, ApJ, 483, 95L 
\bibitem[Sugizaki et al. 2001]{sugizaki01}Sugizaki M., Mitsuda K., Kaneda H. et al. 2001, ApJS, 134, 77 
\bibitem[Tanaka et al. 1999]{tanaka99}Tanaka Y., Miyaji T, Hasinger G. 1999, Astron. Nahr. 320, 181 
\bibitem[Tanaka 2000]{tanaka00} Tanaka Y. 2000, PASJ, 52, 25L
\bibitem[Tanuma et al. 1999]{tanuma99} Tanuma S., Yokoyama T., Kudoh T. et al. 1999, PASJ, 51, 161
\bibitem[Townes 1989]{townes89} Townes C. 1989, in IAU Simp. 136 
\bibitem[Ueda et al. 1999]{ueda99}Ueda Y., Takahashi T., Inoue H. et al. 1999, ApJ, 518, 656
\bibitem[Valinia \& Marshall 1998]{valinia} Valinia A., Marshall F. 1998, ApJ, 505, 134
\bibitem[Voges et al. 1999]{rass} Voges W., Aschenbach B., Boller Th. et al. 1999, A\&A, 349, 389
\bibitem[Warwick et al. 1985]{warwick} Warwick R., Turner M., Watson M. et al. 1985, Nature, 317, 218
\bibitem[Wood et al. 1984]{wood84}Wood K. S., Meekins J. F., Yentis D. J. et al. 1984, ApJS, 56, 507 
\bibitem[Worrall et al. 1982]{worrall} Worrall D., Marshall F., Boldt E. et al. 1982, ApJ, 255, 111
\bibitem[Yamasaki et al. 1997]{yamasaki} Yamasaki N., Ohashi T., Takahara F. et al. 1997, ApJ, 481, 821
\bibitem[Yamauchi \& Koyama 1993]{}Yamauchi S, Koyama K. 1993, ApJ, 404, 620
\end{thebibliography}
\end{document}